# Multiscale light-matter dynamics in quantum materials: from electrons to topological superlattices

Taufeq Mohammed Razakh
*Collaboratory for Advanced Computing and Simulations*
*University of Southern California*
Los Angeles, CA, USA
razakh@usc.edu

Thomas Linker
*Stanford PULSE Insitute*
*SLAC National Accelerator Laboratory*
Menlo Park, CA, USA
tlinker@slac.stanford.edu

Ye Luo
*Computational Science Division*
*Argonne National Laboratory*
Lemont, IL, USA
yeluo@anl.gov

Nariman Piroozan
*Intel Corporation*
Santa Clara, CA, USA
piroozan@alumni.usc.edu

John Pennycook
*Intel Corporation*
Santa Clara, CA, USA
john.pennycook@intel.com

Nalini Kumar
*Intel Corporation*
Santa Clara, CA, USA
nalini.kumar@intel.com

Albert Musaelian
*School of Engineering and Applied Sciences*
*Harvard University*
Cambridge, MA, USA
amusaelian@alumni.harvard.edu

Anders Johansson
*School of Engineering and Applied Sciences*
*Harvard University*
Cambridge, MA, USA
andersjohansson@g.harvard.edu

Boris Kozinsky
*School of Engineering and Applied Sciences*
*Harvard University*
Cambridge, MA, USA
bkoz@seas.harvard.edu

Rajiv K. Kalia
*Collaboratory for Advanced Computing and Simulations*
*University of Southern California*
Los Angeles, CA, USA
rkalia@usc.edu

Priya Vashishta
*Collaboratory for Advanced Computing and Simulations*
*University of Southern California*
Los Angeles, CA, USA
priyav@usc.edu

Fuyuki Shimojo
*Department of Physics*
*Kumamoto University*
Kumamoto, Japan
shimojo@kumamoto-u.ac.jp

Shinnosuke Hattori
*Advanced Research Laboratory, Research Platform*
*Sony Group Corporation*
Atsugi, Kanagawa 243-0014, Japan
shinnosuke.hattori@gmail.com

Ken-ichi Nomura
*Collaboratory for Advanced Computing and Simulations*
*University of Southern California*
Los Angeles, CA, USA
knomura@usc.edu

Aiichiro Nakano
*Collaboratory for Advanced Computing and Simulations*
*University of Southern California*
Los Angeles, CA, USA
anakano@usc.edu

*Abstract*—Light-matter dynamics in topological quantum materials enables ultralow-power, ultrafast devices. A challenge is simulating multiple field and particle equations for light, electrons, and atoms over vast spatiotemporal scales on Exaflop/s computers with increased heterogeneity and low-precision focus. We present a paradigm shift that solves the multiscale/multiphysics/heterogeneity challenge harnessing hardware heterogeneity and low-precision arithmetic. Divide-conquer-recombine algorithms divide the problem into not only spatial but also physical subproblems of small dynamic ranges and minimal mutual information, which are mapped onto best-characteristics-matching hardware units, while metamodel-space algebra minimizes communication and precision requirements. Using 60,000 GPUs of Aurora, DC-MESH (divide-and-conquer Maxwell-Ehrenfest-surface hopping) and XS-NNQMD (excited-state neural-network quantum molecular dynamics) modules of MLMD (multiscale light-matter dynamics) software were 152- and 3,780-times faster than the state-of-the-art for 15.4 million-electron and 1.23 trillion-atom $PbTiO_3$ material, achieving 1.87 EFLOP/s for the former. This enabled the first study of light-induced switching of topological superlattices for future ferroelectric 'topotronics'.

## I. Justification For Gordon Bell

First end-to-end exa-deployed multiscale light-matter dynamics simulation and machine learning for light, electrons, and atoms, achieving 152× and 3,780× improvements in time-to-solution compared to state-of-the-art for 15.4 million-electron quantum dynamics and 1.23 trillion-atom neural-network molecular dynamics, respectively, with 1.87 EFLOP/s for the former.

## II. Performance Attributes

| Category of achievement | Time-to-solution, peak performance, scalability |
|---|---|
| Performance | $1.11 \times 10^{-7}$ [sec/(electron•step)], 1.87 EFLOP/s, ~100% weak-scaling parallel efficiency for quantum dynamics; and $1.88 \times 10^{-15}$ [sec/(atom • weight • step)], 99.7% weak-scaling parallel efficiency for neural-network molecular dynamics |
| Type of method used | Maxwell, Schrödinger, Newton, and neural-network dynamics |
| Results reported on the basis of | Whole application |
| Precision reported | Mixed precision |
| System scale | Full-scale system: 60,000 GPUs |
| Measurement mechanism | Timers and FLOP count |

## III. Problem Overview: Multiscale/Multiphysics Light-Matter Dynamics at a HPC Crossroads

Light-matter dynamics in quantum materials holds promise for a sustainable society with ubiquitous artificial intelligence (AI). The enormous power required by AI demands ultralow-power and ultrafast computing and sensing devices [1] that are best enabled by quantum materials, in which quantum mechanics (QM), such as the topology of electronic wave functions, essentially governs functionality [2]. While topological quantum matter was the topic of the 2016 Nobel physics prize, ultrafast control of quantum materials on demand is best achieved *via* light-matter interaction [3]. Specifically, nonlinear interaction of laser light with matter generates attosecond ($10^{-18}$ second) pulses. The new era of attosecond physics was heralded by the 2023 Nobel physics prize to Agostini, Krausz, and L'Hullier.

Classically, light dynamics is described by Maxwell's equations published in 1861, whereas molecular dynamics (MD) in matter is described by Newton's equations published in 1687. Quantum description of matter was the hallmark of science in the 20th century, but its application was hampered by the exponential computational complexity. For static quantum properties, the complexity was reduced to $O(N^3)$ ($N$ is the number of electrons) by density functional theory (DFT) [4], for which Walter Kohn received the 1998 Nobel chemistry prize. The complexity was further reduced to $O(N)$ by linear-scaling DFT algorithms [5] based on a physical data-locality principle called quantum nearsightedness [6]. For light-matter dynamics, time-dependent Schrödinger equations in time-dependent density functional theory (TDDFT) need be solved instead for electrons along with Maxwell's equations for light [7, 8].

There are two complementary first-principles approaches to describe quantum dynamics (QD) of electrons coupled with MD of atoms, *i.e.*, **nonadiabatic quantum molecular dynamics (NAQMD)** [9]: (1) Ehrenfest dynamics to describe short-time transient MD for atoms along with QD for electrons as in Maxwell+TDDFT simulations [7, 8]; and (2) surface hopping (SH) to describe longer-time MD driven by electronic transitions. Quantum uncertainty principle separates Ehrenfest and SH time-scales at $t \sim \hbar/\Delta E$ ($\sim 10^{-15}$ s), where $\Delta E$ is the separation between key electronic energy levels that dictate light-induced MD, and $\hbar$ is the Planck constant. While both are indispensable for understanding optical control of materials [10, 11], their integration into single software has remained elusive. Here, we present a new multiscale NAQMD approach within a divide-and-conquer (DC) scheme named **DC-MESH (divide-and-conquer Maxwell-Ehrenfest-surface hopping)** that seamlessly integrates Ehrenfest- and SH-NAQMD across time scales, along with Maxwell's equations for light (Fig. 1) [12].

Light-matter dynamics is a challenging multiscale problem that encompasses the above multiphysics of vastly different computational characteristics at diverse spatiotemporal scales. Its temporal scales extend over fast ($10^{-18}$ s) elementary processes of light-electron coupling and much slower ($10^{-9}$ s) materials response through electron-atom coupling. In addition, disparate length scales need be accounted for, ranging from spatial variations of electronic wave functions ($10^{-11}$ m) to large topological features of quantum materials for optoelectronic device applications ($10^{-6}$ m). Such multiscale 'topotronics' beyond the reach of first-principles NAQMD ($\sim 10^{-12}$ s and $10^{-9}$ m) has computationally been addressed by 'second-principles' approaches, in which approximate classical equations of motion are derived based on first-principles QM calculations [13]. Recent application of machine learning (ML) to **neural-network quantum molecular dynamics (NNQMD)** simulations has revolutionized this area, providing first-principles QM accuracy at a fraction of computational cost by replacing electronic wave functions by deep neural networks [14-16] (ML-enabled MD was cited in the 2024 Nobel physics prize announcement, which heralded the central role of AI in the very core of basic science). We have extended NNQMD to **excited-state NNQMD (XS-NNQMD)** [11], allowing light-induced switching of large topological structures of device relevance to be studied from first principles for the first time (Fig. 1).

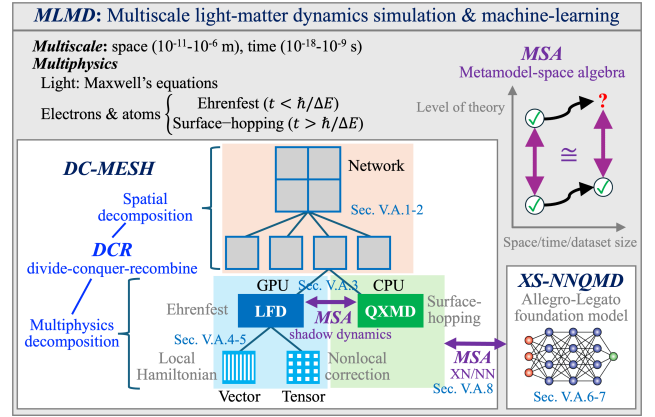

Fig. 1: Software and enabling technologies: **MLMD** (multiscale light-matter dynamics) simulation & machine-learning software integrates **DC-MESH** (divide-and-conquer Maxwell-Ehrenfest-surface hopping) and **XS-NNQMD** (excited-state neural-network quantum molecular dynamics) modules. **DCR** (divide-conquer-recombine) algorithms decompose the problem into not only spatial but also multiphysics subproblems that are mapped onto best characteristics-matching hardware units, while **MSA** (metamodel-space algebra) allows the subproblems to reside in respective hardware units and minimizes inter-unit communications.

Concurrent to these new scientific developments, the computing landscape is changing rapidly. In particular, high performance computing (HPC) nodes are becoming increasingly more heterogeneous by integrating different functional units (or chiplets), largely focusing on low-precision arithmetic accelerators to serve the market-dominating AI applications [17]. Thus, HPC is at a historic crossroads, where traditional modeling and simulation applications may not survive. This paper presents a paradigm shift that solves the multiscale/multiphysics/heterogeneity/low-precision challenge by leveraging hardware heterogeneity and low-precision arithmetic as resources rather than regarding them as obstacles, through innovations explained in Sec. V.

## IV. CURRENT STATE OF THE ART

While high-end supercomputing has successfully been applied to quantum-mechanical study of *static* material properties, including a number of Gordon-Bell prizes [18-20], its application to *quantum dynamics* such as attosecond physics remains in its infancy [21-23]. Several software packages exist for Maxwell+Ehrenfest simulations on parallel computers such as Octopus [24] and SALMON [25], where a multiscale DC approach has been applied to the Maxwell-Ehrenfest (ME) subproblem [25], but not to the whole Maxwell-Ehrenfest-surface hopping (MESH) problem considered in this work (Fig. 1). Also never attempted is the integration of MESH-complete NAQMD with first-principles-accuracy XS-NNQMD as we do in our MLMD application (Fig. 1). As such, we describe below the state of the art (SOTA) for available subsets of MLMD.

The multiscale MESH simulation concept is relatively new [10, 11], with the first integrated MESH program presented only very recently [12]. Accordingly, we here examine the SOTA of the ME subproblem of MESH, for which there exit several high-end computing results in the literature. In [21], Qb@ll code was used to simulate aluminum involving 59,400 electrons, where one QD time step took 53.2 seconds of wall-clock time on 98,304 IBM BlueGene/Q nodes, achieving floating-point performance of 8.75 PFLOP/s (43.5% of theoretical peak). We define the time-to-solution (T2S) of ME simulation as the wall-clock time per QD time step divided by the number of simulated electrons. The T2S of the Qb@ll code for this run is 53.2 [sec]/59,400 [electrons] = $8.96 \times 10^{-4}$ [sec] per electron. In [22], PWDFT code was used to simulate 3,072 spin-degenerate electronic wave functions in silicon, where one QD time step took 260.9 seconds of wall-clock time on 768 GPUs of Summit supercomputer. While this amounts to a raw T2S of $8.49 \times 10^{-2}$ [sec], their parallel-transport time-integration scheme allows the use of 100-times larger time step compared to conventional integrators, effectively reducing the computing time by a factor of 100. The effective T2S is thus $8.49 \times 10^{-4}$ [sec]. More recently, SALMON code was used to simulate 71,040 electrons in 1.2 seconds using 27,648 nodes of Fugaku supercomputer, achieving a T2S of $1.69 \times 10^{-5}$ [sec] [23]. Table I compares these SOTA performance numbers for ME-NAQMD with our result for the MESH-NAQMD superset.

Table I: State-of-the-art Maxwell-Ehrenfest simulations (subset of the Maxwell-Ehrenfest-surface hopping as in DC-MESH of this work).

| Work | Benchmark system | Machine | Time-to-solution [sec] | PFLOP/s (% of FP64 peak) |
|---|---|---|---|---|
| Qb@ll (2016) [21] | Aluminum, 59,400 electrons | IBM BlueGene/Q | $8.96 \times 10^{-4}$ | 8.75 (43.5) |
| PWDFT (2020) [22] | Silicon, 3,072 electrons | Summit | $8.49 \times 10^{-4}$ | 0.12 (2.0) |
| SALMON (2022) [23] | Silica, 71,040 electrons | Fugaku | $1.69 \times 10^{-5}$ | 2.69 (3.17) |
| This work | PbTiO$_3$, 15,360,000 electrons | Aurora | $1.11 \times 10^{-7}$ | 1873 (100.2) |

Also unexplored is multiscale handshaking of NAQMD and *excited-state (XS)* NNQMD to enable seamless simulations from electronic- and atomistic- to device-scales, though *ground-state (GS)* NNQMD has been implemented on high-end supercomputers [15]. In this work, we employ the Allegro NNQMD model, which achieved the SOTA equivariant deep-learning accuracy with 1,000-fold improvement in T2S over previous quantum simulation and improved sample efficiency over the previous SOTA in the GS-NNQMD setting [26]. The robustness of the Allegro model was further improved by the Allegro-Legato model [27], while its generalizability was extended to diverse material properties and processes by the Allegro foundation model (FM) [28]. This work is the first to apply these new developments to XS-NNQMD. The only XS-NNQMD speed reported to date used a much smaller and less accurate neural-network model compared to the equivariant-accuracy Allegro [11]. To account for the accuracy-speed trade-off, we define the T2S of XS-NNQMD as the wall-clock time per MD time step divided by the product of the number of atoms and that of neural-network weights. In [11], one MD step to simulate 1,007,271,936,000-atom PbTiO$_3$ material using 440-weight XS-NNQMD took 3,142.66 seconds on Intel Theta computer. This amounts to T2S of 3,142.66 [sec]/ (1,007,271,936,000 [atoms] × 440 [weights]) = $7.091 \times 10^{-12}$ [sec]; see Table II.

Table II. State-of-the-art XS-NNQMD simulations.

| Work | Machine | Time-to-solution [sec] |
|---|---|---|
| Linker *et al*. (2022) [11] | Theta | $7.09 \times 10^{-12}$ |
| This work | Aurora | $1.88 \times 10^{-15}$ |

## V. INNOVATIONS REALIZED: METASCALABLE MULTISCALE/MULTIPHYSICS PARADIGM

We present a metascalable (*i.e.*, design-once, scale on future architectures [29]) paradigm that solves the multiscale/multiphysics/heterogeneity/low-precision challenge posed in Sec. III by harnessing hardware heterogeneity and low-precision arithmetic. We apply the paradigm to a **multiscale light-matter dynamics (MLMD)** simulation and machine-learning software that seamlessly integrates first-principles **DC-MESH** (divide-and-conquer Maxwell-Ehrenfest-surface hopping) and AI-enhanced **XS-NNQMD** (excited-state neural-network quantum molecular dynamics) modules. This innovation has for the first time allowed end-to-end simulations of optically controlled topotronics extending from the atomistic to device scales in a hardware-optimal manner on an Exaflop/s computer (Fig. 1).

The *Algorithmic innovations* in this work are:
(1) **Divide-conquer-recombine (DCR) algorithms** divide a multiscale/multiphysics problem into not only spatial (Sec. V.A.1) but also physical (Sec. V.A.3-5) subproblems of different computational characteristics, which are separately solved using appropriate computational methods on best-matching hardware units before recombined into a total solution [12, 30].
(2) **Metamodel-space algebra (MSA)** allows the key data structures of subproblems to reside in respective hardware units, while minimizing inter-unit communications. MSA integrates multiple methods using arithmetic operations in a metamodel space, where one axis is the level of theory and the other is the space (Sec. V.A.8 [31-33])/time (Sec. V.A.3 [12])/dataset (Sec. V.A.7 [28]) size. DCR/MSA paradigm delineates multiphysics subproblems, each with a small dynamic range, which in turn maps well onto AI hardware accelerators with various precisions (Sec. V.B.7) [34].
(3) **Globally scalable and locally fast (GSLF) [30]—or globally sparse yet locally dense (GSLD) [35]—solvers** implement DCR algorithms efficiently on a network of GPU-accelerated computing nodes (Sec. V.A.2).
(4) **Allegro-FM**: Accurate, fast, robust, and foundational machine learning model describes diverse materials downstream tasks, based on group-theoretical equivariance [36], local descriptors [36], sharpness-aware training [27], and MSA to unify multifidelity training databases [28] (Sec. V.A.6-7).

*Implementation innovations* complement the algorithmic innovations for exascale performance optimization: (1) Open programming approach based on OpenMP target for portability across exascale computing platforms; (2) data/loop reordering; (3) blocking/tiling; (4) hierarchical parallelization; (5) converting nonlocal correction into dense matrix multiplications ('GEMMification'); (6) GPU-resident kernels; (7) parameterized mixed-precision computation; (8) ahead-of-time compilation; and (9) block-model inference.

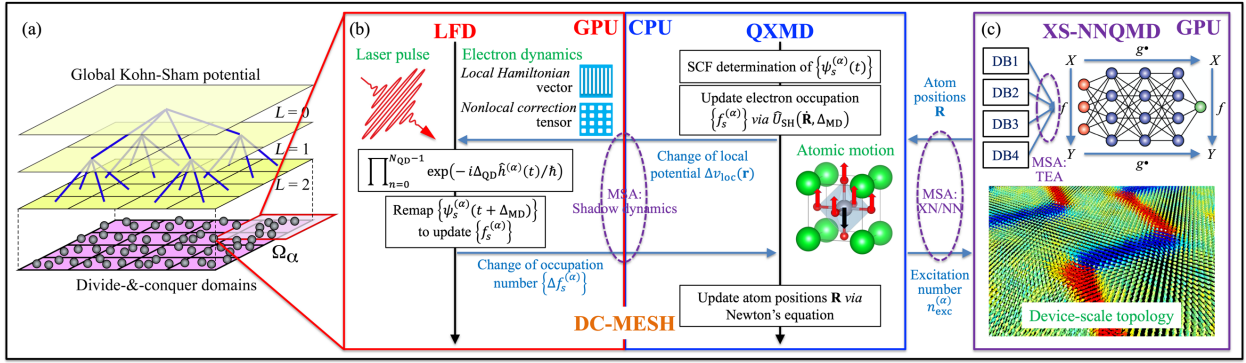

Fig. 2: (a) Divide-and-conquer domains embedded in a global potential. (b) DC-MESH module consists of (i) LFD to describe light-electron interaction on GPU and (ii) QXMD to describe electron-atom coupling on CPU, with minimal CPU-GPU data transfer *via* metamodel space algebra (MSA). (c) Machine learning-based XS-NNQMD extends light-matter dynamics to large device-level spatiotemporal scales.

*A. Algorithmic Innovations*

*A.1 Spatial divide-and-conquer (DC)–DCR1*. The first level of DCR is spatial decomposition (Fig. 1). DFT reduces the exponential complexity of the quantum many-body problem to $O(N^3)$ by self-consistently solving $N$ one-electron problems instead of directly solving the intractable $N$-electron problem [4]. Among various $O(N)$ DFT approaches [5], we employ the DC-DFT algorithm, in which the three-dimensional space $\Omega$ is decomposed into spatially localized domains, $\Omega = \cup_\alpha \Omega_\alpha$ (Fig. 2a), and local electronic Kohn-Sham (KS) wave functions within the domains and the global KS potential are determined by global-local self-consistent-field (SCF) iterations [37]. DC-MESH adopts hierarchical MPI parallelization by assigning one MPI communicator per domain, each handled by multiple MPI ranks through hybrid band-space decomposition, which subdivides KS orbitals (or bands) or space among ranks, depending on a specific computational task [30].

*A.2 Globally scalable and locally fast (GSLF) [30]—or globally sparse yet locally dense (GSLD) [35]—solvers*. We combine: (1) $O(N)$ tree-based multigrid method, which is sparse and scalable, to represent global KS potential; and (2) fast Fourier transform (FFT) to represent local KS wave functions (Fig. 2a) [30]. These solvers apply to the mean-field electrostatic contribution to the KS potential (*i.e.*, Hartree potential). On the other hand, nonlocal exchange-correlation (xc) and nonlocal pseudopotential effects, which represent higher-order correlations and complex chemical interactions, act on the entire spatial extent of each KS wave function at once [38]. Such dense computations are performed only within each domain, taking advantage of their short spatial ranges based on the quantum-nearsightedness locality principle (Fig. 1) [6].

*A.3 Shadow dynamics–DCR2/MSA1*. This is the second level of DCR along with the first type of MSA (Figs. 1 and 2b). The key insight is that fundamental physics equations are all local at the finest spatiotemporal scales, *i.e.*, simple partial differential equations with differential operators acting locally. On the other hand, coarse-grained schemes to approximately describe complex chemical interactions come with an excessive computational cost of nonlocal operations. Simple data parallelism in the former—which we call **Local Field Dynamics (LFD)**—fits naturally to hardware accelerators such as GPU (Figs. 1 and Fig. 2b). On the other hand, complex chemical interaction in the latter—which we call **Quantum eXcitation Molecular Dynamics (QXMD)**—can take advantage of complex instruction sets in CPU (Figs. 1 and 2b). To minimize data transfer between CPU and GPU, we adopt a shadow dynamics approach [39], in which a GPU-resident proxy is solved to capture effective action of LFD on QXMD through electronic occupation numbers, $f_s^{(\alpha)} \in [0,1]$ (for the $s$-th KS wave function in the $\alpha$-th domain) [10, 11], which are negligible compared to the large memory footprint of KS wave functions $\psi_s^{(\alpha)}(\mathbf{r})$ represented on many spatial grid points.

*A.4 Local-nonlocal split-operator LFD–DCR3*. The third level of DCR is multiphysics decomposition between local Hamiltonian dynamics and nonlocal correction within LFD, which can be formulated as vector and tensor computations, respectively (Figs. 1 and 2b); see Sec. A5. In MD simulation, time evolution of $N_{\text{atom}}$ atoms is achieved by repeatedly updating $3N_{\text{atom}}$-element position and velocity vectors, $\mathbf{R}$ and $\dot{\mathbf{R}}$, using a time step of $\Delta_{\text{MD}} \sim 10^3$ attoseconds:
$$\big(\mathbf{R}(t+\Delta_{\text{MD}}), \dot{\mathbf{R}}(t+\Delta_{\text{MD}})\big) = \exp(\hat{L}\Delta_{\text{MD}})\big(\mathbf{R}(t), \dot{\mathbf{R}}(t)\big), \quad (1)$$
where $\hat{L}$ is the Liouville operator in classical mechanics. In DC-MESH, faster time evolution of electronic wave functions during each MD step is achieved by

$$\big|\psi_s^{(\alpha)}(t+\Delta_{\text{MD}})\big\rangle =$$
$$\prod_{n=0}^{N_{\text{QD}}-1}\left[\frac{1-\frac{i\Delta_{\text{QD}}}{2\hbar}\hat{v}_{\text{nl}}^{(\alpha)}}{\left\|1-\frac{i\Delta_{\text{QD}}}{2\hbar}\hat{v}_{\text{nl}}^{(\alpha)}\right\|}\exp\left(-\frac{i\Delta_{\text{QD}}}{\hbar}\hat{h}_{\text{loc}}^{(\alpha)}\left(t+\left(n+\frac{1}{2}\right)\Delta_{\text{QD}}\right)\right)\frac{1-\frac{i\Delta_{\text{QD}}}{2\hbar}\hat{v}_{\text{nl}}^{(\alpha)}}{\left\|1-\frac{i\Delta_{\text{QD}}}{2\hbar}\hat{v}_{\text{nl}}^{(\alpha)}\right\|}\right]\hat{U}_{\text{SH}}\big(\dot{\mathbf{R}},\Delta_{\text{MD}}\big)\big|\psi_s^{(\alpha)}(t)\big\rangle, \quad (2)$$

where $\big|\psi_s^{(\alpha)}(t)\big\rangle$ is the $s$-th complex-valued KS wave function within DC domain $\Omega_\alpha$ at time $t$, $N_{\text{QD}} = \Delta_{\text{MD}}/\Delta_{\text{QD}}$ is the number of QD time steps per MD step ($\Delta_{\text{QD}} \sim 1$ attosecond), $\hat{U}_{\text{SH}}\big(\dot{\mathbf{R}},\Delta_{\text{MD}}\big)$ is the surface-hopping procedure to update the electron occupation $f_s^{(\alpha)}$ perturbatively according to nonadiabatic coupling arising from slow atomic motions [9], $i = \sqrt{-1}$, and the electronic Hamiltonian operator is

$$\hat{h}^{(\alpha)} = \frac{1}{2m}\left(\frac{\hbar}{i}\frac{\partial}{\partial \mathbf{r}} + \frac{e}{c}\mathbf{A}_{\mathbf{X}(\alpha)}(t)\right)^2 + v_{\text{loc}}^{(\alpha)}(\mathbf{r},\mathbf{R},t) + \hat{v}_{\text{nl}}^{(\alpha)} = \hat{h}_{\text{loc}}^{(\alpha)}(t) + \hat{v}_{\text{nl}}^{(\alpha)}. \quad (3)$$

Here, $m$ and $e$ are the electron mass and charge, $\hbar$ is the Planck constant, $c$ is the light speed, and $\mathbf{A}_{\mathbf{X}(\alpha)}$ is the electromagnetic vector potential at the spatial position $\mathbf{X}(\alpha)$ of the α-th domain, which is determined by solving Maxwell's equations. In Eq. (3), the local potential $v_{\text{loc}}^{(\alpha)}$ operates spatial-point-by-point, whereas the nonlocal operator $\hat{v}_{\text{nl}}^{(\alpha)}$ here collectively denotes the nonlocal ionic pseudopotential and nonlocal exchange-correlation potential, which has much more complex computational characteristics [38]. In the shadow dynamics approach in Sec. A3, only a small change $\Delta v_{\text{loc}}^{(\alpha)}$ in local potential due to slight atomic movements during $\Delta_{\text{MD}}$ is passed from QXMD to LFD (Fig. 2b). In return, LFD passes back the change of electron occupation $f_s^{(\alpha)}$ due to light-matter coupling (Fig. 2b). These CPU-GPU data transfers are amortized by subsequent $N_{\text{QD}}$ (~$10^3$) QD time steps without any data transfer.

*A.5 Local-vector and nonlocal-tensor solvers.* To expose natural data-parallelism in local Hamiltonian dynamics, $\exp(-i\Delta_{\text{QD}}\hat{h}_{\text{loc}}^{(\alpha)}/\hbar)$, in Eq. (2), we employ finite-difference representation of wave functions and adopt **data-parallel local Hamiltonian-dynamics solvers** that perform uniform operations on nearest-neighbor mesh points [40]: (1) a block-diagonal split-operator solver for electronic wave functions [41]; and (2) an iterative dynamical simulated annealing (DSA) solver for the Hartree potential [42]. To ensure stable time propagation, we employ a self-consistent, time-reversible unitary approach that handles nonlinearity, *i.e.*, the time-propagation operator itself depends on the wave functions being propagated [43]. On the other hand, we apply nonlocal $\hat{v}_{\text{nl}}^{(\alpha)}$ in Eq. (2) instead in a vector space spanned by KS wave functions [44]. This switch of representation **converts nonlocal correction to dense matrix multiplications** [12], as will be detailed in Sec. B.5.

*A.6 Allegro-Legato: fast and robust XS-NNQMD.* The next algorithmic innovation deals with the XS-NNQMD module of the MLMD software (Fig. 2c). A recent breakthrough in NNQMD has drastically improved the accuracy of inter-atomic force prediction over previous models, which was achieved *via* rotationally equivariant neural networks based on a group theoretical formulation of tensor fields [45]. SOTA accuracy has now been combined with a record speed based on spatially localized descriptors in the latest NNQMD model named Allegro (meaning fast), making it the first exa-deployable model with the SOTA equivariant accuracy [36].

Despite its remarkable computational scalability, exascale NNQMD simulations face a major unsolved issue known as *fidelity scaling*, *i.e.*, small prediction errors propagate and lead to unphysical atomic forces that even cause the simulation to terminate unexpectedly [27]. As simulations become spatially larger and temporarily longer, the number of unphysical force predictions increases proportionally, which severely limits the fidelity of exascale NNQMD simulations, especially for far-from-equilibrium XS-NNQMD. We reduce the number of unphysical force-prediction outliers by enhancing the robustness of the model through sharpness-aware minimization (SAM) [46], *i.e.*, regularizing the curvature of the loss surface during training. The resulting Allegro-Legato (meaning fast and "smooth") model elongates the time-to-failure $t_{\text{failure}}$, while maintaining the same inference speed and accuracy [27]. Allegro-Legato exhibits much weaker dependence of time-to-failure on the problem size, $t_{\text{failure}} \propto N_{\text{atom}}^{-0.14}$ compared to the Allegro model ($t_{\text{failure}} \propto N_{\text{atom}}^{-0.29}$) [27]. This breakthrough has enabled spectroscopically-stable, long-time MD simulations for the first time to reproduce the fine vibrational structures observed in SOTA neutron-scattering experiments [47].

*A.7 Allegro-FM: universal NNQMD–MSA2.* Foundation models (FM) are a paradigm shift in AI, where a single universal model acquires sufficient generalizability to enable diverse, out-of-distribution downstream tasks [48]. Our equivariant Allegro-FM describes a wide variety of material properties and processes accurately with a single pretrained model, covering 89 elements in the periodic table, and is applicable to diverse downstream tasks including structural correlations, reaction kinetics, mechanical strengths, fracture, and solid/liquid dissolution, exhibiting emergent capabilities for which the model was not trained [28]. This universality arises from a large training database unifying multiple first-principles training datasets with varying fidelities (*e.g.*, different exchange-correlation functionals). This unification in turn is achieved by a **total energy alignment (TEA)** framework [49], which is the second type of MSA that uses affine (shift and scale) transformations in a metamodel space (Fig. 2c).

*A.8 Multiscale XN/NN–MSA3.* Minimal-communication handshaking between DC-MESH and XS-NNQMD is achieved by the third type of MSA (Figs. 1 and 2) akin to multiscale quantum-mechanics/molecular-mechanics (QM/MM) method, which was the topic of the 2013 Nobel chemistry prize to Karplus, Levitt, and Warshel. QM/MM can be formulated as extrapolation in a metamodel space, where the two axes are model accuracy and problem size [31, 32, 50]. The sole assumption is that the difference between QM and MM methods remains the same across problem sizes (Fig. 1). We have applied this MSA to adaptive multiscale QM/MM simulations, in which compute-intensive QM computations are dynamically embedded in lower-fidelity MM computation only where and when high fidelity is called for [51]. We have recently extended the MSA to multiscale NN/MM, which embeds first-principles-accuracy NNQMD in MM [33]. In this paper, we further extend the MSA to XN/NN, where excited-state XS-NNQMD is perturbatively added to ground-state GS-NNQMD. Here, GS-NNQMD is based on the pretrained Allegro-FM in Sec. A7, which is fine-tuned with additional NAQMD training data to generate an XS-NNQMD model for describing photoexcitation.

XS-NNQMD passes atomic positions to DC-MESH, whereas DC-MESH returns a number of photoexcited electrons $n_{\text{exc}}^{(\alpha)}$ per domain α (which is computed from $f_s^{(\alpha)}$) to XS-NNQMD (Fig. 2, b and c). We collect $n_{\text{exc}}^{(\alpha)}$ using MPI gather across domains only once at the end of DC-MESH, thus preventing frequent pauses to GPU activity. In each MD step, GS- and XS-NNQMD models independently predict atomic force on the $i$-th atom $\mathbf{F}_i$ as $\mathbf{F}_{GS}^i$ and $\mathbf{F}_{XS}^i$ based on the same tensor object inputs, then the predicted forces are combined as

$$\mathbf{F}_i = (1-w)\mathbf{F}_{GS}^i + w\mathbf{F}_{XS}^i, \quad (4)$$

where $w$ is the fraction of XS model. The value of $w$ is determined by the electronic excitation number $n_{exc}^{(\alpha)}$ [11]. To minimize costly data movement, all models per MD domain are offloaded to GPU at the beginning of simulation and the force inference from these models is done on single GPU tile.

*B. Implementation Innovations*

In this section, we describe a series of optimizations applied to improve the performance of MLMD on Exaflop/s platforms.

*B.1 Open programming approach*. For portability across supercomputers in the US and elsewhere, we use open programming approaches for parallel programming: Message Passing Interface (MPI) for message passing and Open Multi-Processing (OpenMP) for multithreading. To facilitate minimally invasive offloading to GPU, we use high-level OpenMP target constructs (except for the use of low-level SYCL only through BLAS library) instead of proprietary languages. We also avoid unnecessary overheads by creating a common device data environment to reduce the overall amount spent in host-to-device data transfer in the OpenMP target region. In addition to these open parallel-programming approaches, machine-learning (ML) modules in XS-NNQMD use the most widely used PyTorch. Our open parallel and ML programming approach not only achieves EFLOP/s performance for a specific application on a particular platform but also brings it to general ML-integrated high-performance scientific computing on any computing platform around the globe.

*B.2 Data and loop re-ordering*. Data-parallel solvers in Sec. A.5 introduce sparse stencil operations with strided data access in the $x$, $y$, and $z$ directions. To achieve optimal memory access patterns in registers, we make the array of wave functions for $N_{orb}$ KS orbitals each on $N_{grid}$ spatial grid points as a structure of arrays (SoA), *i.e.*, consecutively store the complex values for $N_{orb}$ orbitals for each grid point. Accordingly, wave-function-update loops are ordered such that the fastest-changing index corresponds to the orbital. This allows space-dependent stencil operators to be reused for $N_{orb}$ orbitals on both CPU and GPU [12].

*B.3 Blocking/tiling*. We further block each innermost loop over $N_{orb}$ orbitals, so that the wave functions accessed in the loop fit into the cache. The added loop of blocks also allows distributing the computation to more GPU blocks when offloading is used [12].

*B.4 Hierarchical parallel regions*. Through our loop re-ordering and SoA optimizations, we expose the kernel to a high level of parallelism. For example, propagation of wave functions on the grid points of a $y$-$z$ grid plane can be concurrently computed for an $x$-direction stencil. Hence, the first level of parallelism is achieved as time-evolution requires only knowledge of the wave function at the current time step and the previous step within the same plane. A second level of parallelism comes into effect from the ability to propagate the wave function independently of the orbital. This hierarchical parallelism applies to both single-instruction multiple-data (SIMD) and single-instruction multiple-threads (SIMT) paradigms. The parallelization over planes and orbitals are collapsed into a larger loop. Combined optimizations in Secs. B.2-4 have significantly sped up the computation of local time-propagator $\exp(-i\Delta_{QD}\hat{h}_{loc}^{(\alpha)}/\hbar)$ (Sec. A.5) [12]. Table III shows the reduction of its key kernel, *kin_prop*(), on the Polaris computer at Argonne Leadership Computing Facility, with 2.8 GHz AMD EPYC Milan 7543P CPU and Nvidia A100 GPU. The timing is for 1,000 QD steps involving 64 KS wave functions each on $70 \times 70 \times 72$ finite-difference mesh points. For simplicity, a single GPU timing is compared with a single CPU-core timing. The optimizations B.2 and B.3 have achieved 9.22-fold speedup on CPU, while GPU offloading (B.4) has resulted in 36.1-fold speedup over CPU version. The overall speedup is thus 338-fold over the baseline [12].

Table III. Runtime of the *kin_prop*() function for local time-propagator in the LFD subprogram of the DC-MESH module.

| Implementation | Target | Runtime (s) | Speedup |
|---|---|---|---|
| Baseline | CPU | 8.655 | 1 |
| Data & loop re-ordering (Sec. B.2) | CPU | 2.356 | 3.67 |
| Blocking/tiling (Sec. B.3) | CPU | 0.939 | 9.22 |
| Hierarchical parallel regions (Sec. B.4) | GPU | 0.026 | 338 |

*B.5 GEMMification of nonlocal correction*. The orders-of-magnitude speedup of the local Hamiltonian propagator $\exp(-i\Delta_{QD}\hat{h}_{loc}^{(\alpha)}/\hbar)$ in Table III has left the nonlocal counterpart $\hat{v}_{nl}^{(\alpha)}$ as the performance hotspot for DC-MESH. Switching from finite-difference to KS-orbital representations in Sec. A.5 transforms this computation into matrix multiplications. Let us define a $N_{grid} \times N_{orb}$ wave-function matrix $\Psi(t)$, where $N_{grid}$ and $N_{orb}$ are the number of grid points to represent each wave function and that of KS wave functions. The nonlocal correction in Eq. (2) then reads

$$\Psi(t) -= \delta \Psi(0) \Psi^{\dagger}(0) \Psi(t), \qquad (5)$$

where $\delta$ is a small complex number and $\Psi^{\dagger}$ denotes a Hermitian transpose matrix. We implement Eq. (5) using BLAS (basic linear algebra subprograms) level 3 GEMM (general matrix-matrix multiply) calls [12]. Specifically, we use GEMM kernels on the device by accessing data that is already allocated on GPU with the OpenMP clause *use_device_ptr*(*list*). This allows us to make SYCL-BLAS on an explicitly-defined GPU queue, thus eliminating the need to allocate from host code to take advantage of the high performance oneMKL BLAS. In addition to time-propagation in Eq. (2), GEMMification is applied to nonlocal correction in energy and electric current (within time-dependent current density functional theory, TDCDFT [52]), with the latter used in Maxwell's equations for light.

*B.6 GPU-resident kernels*. The key computational advantage of the shadow dynamics is that the large wave-function arrays, $\Psi(t)$ and $\Psi(0)$, can be made GPU-resident, thereby eliminating massive CPU-GPU data transfer. Such GPU-resident data structures are facilitated by our custom C++ class constructor and destructor based on OpenMP target data constructs. The custom allocator named *OMPallocator* is used for container classes (*e.g.*, *std::vector*), which are intended to be GPU-resident. Upon initialization, the allocator calls *#pragma omp target enter data map(alloc)*, while upon destruction, it calls *#pragma omp target exit data map(delete)*.

This significantly eases the programming burden of managing GPU-resident data, while keeping the use-side code neat. In addition, the *HostAllocator* may be replaced with a customized allocator using pinned host memory to further improve host-device transfer rate [12].

*B.7 Parameterized mixed-precision computation*. In DCR, judicious decomposition of a problem results in subproblems with small dynamic ranges and minimal mutual information. While chemical accuracy for the QXMD subproblem in DC-MESH requires FP64 precision, shadow dynamics requires minimal information from the proxy LFD, *i.e.*, only small change $\Delta f_s^{(\alpha)}$ to occupation number $f_s^{(\alpha)}$, which itself has a limited dynamic range, [0,1]. This can be achieved in FP32 without sacrificing accuracy. Similar decomposition of occupation numbers into mixed-precision arithmetic was adopted in a previous Gordon Bell prize [19]. In DCR, FP64/FP32 decomposition is done instead at a subprogram level by simply using a class template with parameterized precision. Furthermore, nonlocal correction, Eqs. (2) and (5), in LFD is perturbatively added [53] and is constructed to reproduce the dominant energy term exactly [44], which is amenable to even lower-precision GEMM. In fact, brain-floating point 16 (BF16) [54] with FP32 accumulation (hybrid FP32/BF16) was shown to be sufficient for this computation in [34] (see Sec. VI.C). In XS-NNQMD, Allegro-FM uses FP32 to represent normalized weights and internal activations again with limited dynamic range [0,1], while employing FP64 for the final stage of deep learning to compute interatomic forces [26].

*B.8 Ahead-of-time (AOT) compilation*. To enhance code portability, just-in-time (JIT) compilation is used by default on Aurora. However, this incurs large time penalty at the beginning of each simulation. To avoid this, we employ AOT compilation, so that the binary contains the actual assembly code of target platform instead of intermediate SPIR-V code.

*B.9 Block model inference*. GPU memory offers high bandwidth to enable ultrahigh-throughput computation, however, the relatively small storage capacity often becomes a serious bottleneck in the scalability of simulation size. In Allegro model, besides network parameters, atomic position, type, and neighbor-list tensors need be stored in GPU memory. All three tensors scale linearly, though the neighbor-list tensor has a large prefactor, about 50-200, compared to the other two tensors. Thus, we block the model inference calculation in two batches to overcome the limitation in the system scalability and have achieved an order-of-magnitude larger system size compared to the SOTA.

## VI. How Performance Was Measured

### A. Applications Used to Measure Performance

Ultrafast modulation of electronic structures *via* photo-excitation can open hidden pathways to exotic quantum material phases that are not obtainable otherwise [3]. Of particular interest is optical manipulation of emergent polarization topologies in ferroelectric materials [55]. Since toplogical quantum structures like polar skyrmions are characterized by integer topological indices, they are protected from thermal noise [2]. Accordingly, topological optoelectronic devices can be operated at much lower voltages than the current CMOS devices, thereby overcoming the so called Boltzmann's tyranny that imposes practical limits to sustain Moore's law. However, fundamental understanding of topological polarization control under far-from-equllibrium optical exciation remains critically lacking due to the multiscale/multiphysics challenge explained in Sec. III. The open science question is how to control attosecond electronic excitation dynamics initiated by ultrafast laser-light pulses to generate longer-time structural changes relevant for device operation. The performance-optimized MLMD software described in Sec. V has overcome this challenge and has for the first time enabled the study of light-induced topological switching for future ultrafast and ultralow-power ferroelectric topotronics applications from first principles.

We study laser-induced ultrafast dynamics of topological patterns in a prototypical ferroelectric topotronics material, PbTiO$_3$. While an elementary topological pattern like skyrmion may have a length scale of $10^{-8}$ m, studying its device application requires simulation of an array (or superlattice) of such elementary patterns extending $10^{-6}$ m. Such photo-induced topological switching of superlattices has indeed been realized by a recent experimental breakthrough [56]. We adopt a multiscale simulation approach [11], where we first prepare a complex polar topology, *i.e.*, a superlattice of skyrmions using GS-NNQMD. These atomic positions are fed to DC-MESH to simulate electronic and structural responses to a femtosecond laser pulse. Informed by the resulting electronic-excitation number from DC-MESH, XS-NNQMD simulation is then performed to study larger spatiotemporal-scale topological dynamics (Fig. 3).

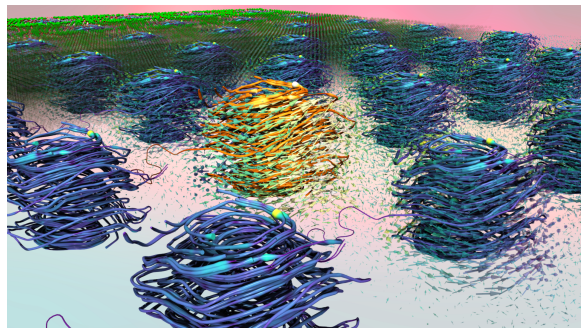

Fig. 3. Photo-switching of a ferroelectric skyrmion superlattice in PbTiO$_3$.

### B. Systems and Environment

We test the performance of our MLMD simulation and machine-learning software on the Aurora supercomputer at Argonne Leadership Computing Facility. It is an HPE Cray EX supercomputer where each node is a vertically installed compute blade, 64 blades per rack, and 166 racks amount to a total of 10,624 blades. Each blade has two 2.00 GHz Intel® Xeon Max 9470 52-core CPUs with 64 GB High Bandwidth Memory (HBM) and 512 GB DDR5 Memory each. Each blade has 6 Intel® Data Center GPU Max 1550 (2 tile) GPU, containing 128 GB of HBM2e memory, and a maximum frequency of 1.6 GHz. The CPU-GPU interconnect is through PCIe and the GPU-GPU interconnect is through Xe Link. The system interconnect is Slingshot 11, which has high radix 64-port switches and offers adaptive routing, congestion control, and bandwidth guarantees by assigning traffic classes to applications and is coupled with Dragonfly topology. PVC tiles per blade have a peak floating-point performance of 187 TFLOP/s, which makes the entire machine's theoretical peak performance ~2 EFLOP/s for FP64.

The MLMD software consist of two modules: DC-MESH and XS-NNQMD. The DC-MESH code consists of QXMD subprogram written in Fortran with MPI and LFD subprogram written in C++ with OpenMP. For performance evaluation on Aurora, DC-MESH is built using Intel® oneAPI 2025.0 release compilers as well as Intel® Math Kernel Library (MKL) 2025.0. Similarly, XS-NNQMD consists of an MD engine written in Fortran with MPI and a C++ code for Allegro model inference. XS-NNQMD is built with Intel® Extension for PyTorch (IPEX) version 2.1.40 and Intel® oneAPI version 2024.2.1.

Floating-point performance of DC-MESH is quantified by measuring total FLOP/s (floating point operations per second). This metric is determined using several software packages for both CPU and GPU. First, Intel® Software Development Emulator (Intel® SDE) is used to determine the total FLOP count of the workload and its breakdown into CPU and GPU. We then measure the wall-clock time on the GPUBLAS with unitrace, part of the Profiling Tools Interfaces for GPU. Dividing the GPU FLOP count by GPU wall-clock time gives us GPU FLOP/s. The same computation is performed on the CPU side to give us CPU FLOP/s. Finally, sum of FLOP/s for CPU and GPU provides the total FLOP/s.

## C. Parametereized Mixed-Precision Implementation

Intel® MKL supports multiple compute modes for BLAS level 3 routines to handle accuracy-speed trade-off. In the float_to_{BF16,BF16x2,BF16x3} modes, the library internally converts single-precision input data to sums of 1, 2, or 3 BF16 values, then uses the fast systolic arrays available on recent discrete GPUs to multiply the resulting BF16 component matrices and accumulate in single precision. As the number of BF16 components increases, so does expected accuracy, to the point that BF16x3 accuracy is comparable to standard single-precision arithmetic. As explained in Sec. V.B.7, float_to_BF16 was found to provide sufficient accuracy for nonlocal correction in DC-MESH in [34].

## VII. Performance Results

We measure the scalability, FLOP/s performance, and T2S of the DC-MESH and XS-NNQMD modules of the MLMD software on the Aurora supercomputer.

### A. Weak and Strong Scalability

*A.1 DC-MESH.* We first perform a weak-scaling benchmark of DC-MESH, in which the number of electrons per MPI rank (or spatial DC domain), $N/P$ is kept constant. Each MPI rank represents $PbTiO_3$ material containing up to 1,024 KS wave functions represented by the plane-wave basis in the QXMD subprogram, while each complex-valued KS wave function in the LFD subprogram is represented on finite-difference mesh points. We use up to 10,000 computing nodes with 12 MPI ranks per node, where each rank is accelerated by one tile of GPU. Each spatial DC domain consists of a mutually exclusive core surrounded by a buffer layer (Fig. 2a). For a buffer thickness equal to half the core domain length in each Cartesian direction, the total problem size excluding the overlap is $(1 + 2 \times 1/2)^3 = 8$ times smaller than the product of the number of electrons in each overlapping domain and that of domains. The largest system on 10,000 nodes with 1,024 KS wave functions per domain thus contains $1,024/8 \times 12 \times 10,000 = 15,360,000$ electrons.

We measure the wall-clock time per MD step, which includes 1,000 QD steps, with scaled workloads — $32P$ and $128P$-electron $PbTiO_3$ material using $P$ MPI ranks (Fig. 4a). By increasing the number of electrons linearly with the number of MPI ranks, the wall-clock time remains nearly constant, indicating excellent weak scalability. To quantify the weak-scaling parallel efficiency, we first define the speed of the DC-MESH program as the product of the total number of electrons and the number of MD simulation steps executed per second. The isogranular speedup is given by the ratio between the speed on $P$ MPI ranks and that on the smallest run ($P = 6,144$) as a reference. The weak-scaling parallel efficiency is the isogranular speedup divided by $P/6,144$. With the granularity of 128 electrons per MPI rank, the parallel efficiency is perfect 1.0 within measurement fluctuation on $P = 120,000$ for 15,360,000-electron $PbTiO_3$ material. This result demonstrates very high scalability of DC-MESH, mainly due to the globally-sparse yet locally-dense electronic solvers [30, 35] within the divide-conquer-recombine algorithmic framework.

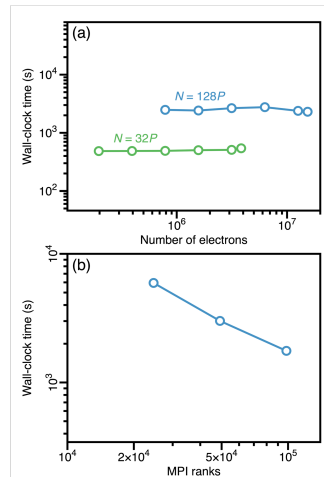

Fig. 4. (a) Weak-scalability of the DC-MESH module with scaled workloads — $32P$ and $128P$-electron $PbTiO_3$ material with $P$ MPI ranks ($P$ = 6,144, ..., 120,000). (b) Strong-scalability of the DC-MESH module as a function of the number of MPI ranks for 12.6M-electron $PbTiO_3$ material.

Next, we perform a strong-scaling test for 12,582,912-electron $PbTiO_3$. In this test, the number of MPI ranks ranges from $P$ = 24,576 to 98,304, while keeping the total problem size constant. Figure 4b shows the wall-clock time per MD simulation step as a function of $P$. The strong-scaling speedup is defined as the wall-clock time on the smallest number of MPI ranks divided by that on a larger number of MPI ranks. The strong-scaling parallel efficiency is the speedup divided by the ratio of the numbers of MPI ranks. It is 0.843 with 98,304 MPI ranks for 12,582,912 electrons. It is more difficult to achieve high strong-scaling parallel efficiency compared with weak-scaling due to the increased communication/computation ratio as the workload reduces.

*A.2 XS-NNQMD.* We also perform weak and strong scalability tests of XS-NNQMD up to 120,000 MPI ranks on 10,000 nodes and 73,800 MPI ranks on 6,150 nodes, respectively. The training dataset of $PbTiO_3$ material was created using Materials Project Trajectory [57] and SPICE [58] datasets combined with TEA framework [49] with the cutoff distance of 5.2 Å. We measure the wall-clock time per MD step averaged over 5 MD steps. Figure 5a shows weak-scaling performance with three different granularities of $N_{atom}/P$ = 160,000, 640,000, and 10,240,000 atoms per MPI rank. The wall-clock time remains nearly constant as a function of the number of MPI ranks. We have achieved an excellent weak-scaling parallel efficiency of 0.997 for the largest granularity of $N_{atom}/P$ = 10,240,000. With smaller granularities, $N_{atom}/P$ = 160,000 and 640,000, respective efficiencies, 0.957 and 0.964, are still excellent despite increased communication/computation ratio. Figure 5b shows strong-scaling performance, where the number of atoms is kept constant. We consider two problem sizes: $N_{atom}$ = 221,400,000 and 984,000,000. We have obtained a decent strong-scaling parallel efficiency of 0.773 for the larger problem size of $N_{atom}$ = 984,000,000. However, the efficiency drops to 0.440 with the smaller problem size $N_{atom}$ = 221,400,000, indicating an excessive communication-to-computation ratio.

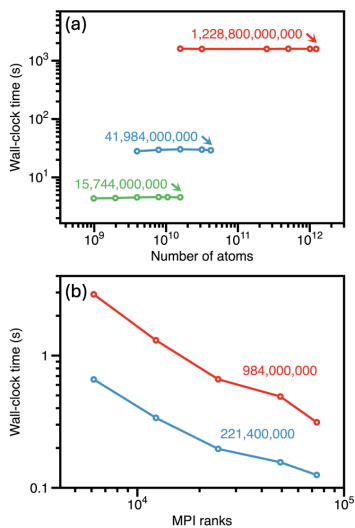

Fig. 5. (a) Weak-scalability of the XS-NNQMD module with 160,000 (green), 640,000 (blue), and 10,240,000 (red) atoms per MPI rank. (b) Strong-scalability of the XS-NNQMD module with problem sizes of 221,400,000 (blue) and 984,000,000 (red) atoms.

*B. Sustained Performance*

We first measure GPU performance of DC-MESH on a single PVC tile by collecting kernel metrics using unitrace. This allows us to analyze the Level Zero, SYCL and OpenMP software layers within the region of interest. Table IV compares TFLOP/s performance, along with its percentage of the theoretical peak FP64 FLOP/s of PVC, for several problem sizes in terms of the number of electrons for $PbTiO_3$. We observe increased FLOP/s performance for larger problem sizes, reaching 17.95 TFLOP/s (78.03% of the peak FP64 performance) using a hybrid FP32/BF16 configuration for precisions (see Sec. VI.C).

Table IV. FLOP/s performance of DC-MESH for several problem sizes on single PVC tile.

| Number of KS orbitals | TFLOP/s | % of FP64 peak (23 TFLOP/s) |
|---|---|---|
| 256 | 5.22 (FP32) | 22.69 |
| 864 | 9.74 (FP32) | 42.35 |
| 1024 | 14.98 (FP32) | 65.16 |
| 1024 | 17.95 (FP32/BF16) | 78.03 |
| 1024 | 7.69 (FP64) | 33.43 |

The increased performance is largely due to the increased arithmetic intensity in the complex GEMM (CGEMM) operations in nonlocal time-propagation. To confirm it, Table V compares TFLOP/s performance of the *nlp_prop*() function for nonlocal time-propagation and the *kin_prop*() function for local time-propagation for the 1,024-orbital run using FP32 in Table IV. Table V also shows TFLOP/s performance of the two CGEMM calls from *nlp_prop*(): one to compute an overlap matrix between KS orbitals at time $t$ and 0, $\Psi(t)$ and $\Psi(0)$, and the other to add a nonlocal correction to $\Psi(t)$ (Eq. 5). The two CGEMM calls with different row-column size combinations indeed achieve 18.72 to 21.66 TFLOP/s (81.39% to 94.17% of the peak performance), which leads to 16.02 TFLOP/s (69.65% of the peak) for the overall performance of *nlp_prop*(). On the other hand, the *kin_prop*() function for local time-propagation achieves 3.51 TFLOP/s (15.26% of the peak). The relatively low performance of the local stencil computation for local time-propagation is consistent with, but much higher than, those in the previous SOTA quantum dynamics, *e.g.*, 2.0% and 3.17% of the peak performance on Summit [22] and Fugaku [23] supercomputers, respectively, as shown in Table I. In more recent work, 7-point star stencil (on which local computation is based) achieves ~3% of the peak performance of the GPU [59], which is again consistent with the previous SOTA values but lower than our result presented in Table V.

Table V. FLOP/s performance of hotspot kernels on single PVC tile for the 1,024-orbital problem in Table IV.

| Kernels | TFLOP/s | % of peak |
|---|---|---|
| CGEMM (1) | 18.72 (FP32) | 81.39 |
| CGEMM (2) | 21.66 (FP32) | 94.17 |
| *nlp_prop*() | 16.02 (FP32) | 69.65 |
| *kin_prop*() | 3.51 (FP32) | 15.26 |

As described in Sec. V.B.7, shadow dynamics allows LFD computation of incremental low-dynamic-range quantities in FP32, while keeping FP64 for computation of complex chemistry in QXMD. While the peak performance of PVC is identical for FP64 and FP32 due to the dual-issued pipes for the former [60], we do observe higher performance for FP32 compared to FP64. For the 1,024-orbital case, Table IV shows 14.98 TFLOP/s (FP32) *vs.* 7.69 TFLOP/s (FP64). On Aurora, peak FP64 performance is restricted to 11 TFLOP/s due to power throttling. FP32 trades precision for a reduced weight value signified with less digits. Less digits results in reduced memory consumption, which in turn improves speed. In Table

IV, we also include results for hybrid FP32/BF16 computation. It uses the BF16 BLAS precision model with FP32 accumulation, reaching 17.95 TFLOP/s (19.83% improvement over FP32) with negligible loss of accuracy (see Sec. V.B.7).

Because of the divide-and-conquer approach, the number of floating-point operations of a total DC-MESH application on multiple computing nodes can be counted by multiplying the number of domains to the above FLOP count obtained from a single domain measurement. FLOP/s performance is then given by the number of aggregated FLOP count divided by the wall-clock time of the entire application. For a 15,360,000-electron problem ($1,024/8 \times 12 \times 10,000$) solved on 10,000 Aurora nodes, the measured performance is 1.873 EFLOP/s, which is 100.2% of FP64 peak or 70.1% of FP32 peak.

### C. Time to Solution

*C.1 DC-MESH*. For 15,360,000-electron problem, one QD step took 1.705 sec on 10,000 Aurora nodes. This amounts to T2S of 1.705 [sec]/15,360,000 [electrons] = $1.11 \times 10^{-7}$ [sec] per electron for the ME-NAQMD subproblem, which is 152-fold reduction of the SOTA [23]; see Table I.

*C.2 XS-NNQMD*. Our XS-NNQMD program uses the Allegro model, which has already achieved 1,000-fold improvement in T2S over previous quantum simulation and improved sample efficiency over the previous SOTA in *ground-state* NNQMD [26]. In this work, we apply the Allegro-FM with improved fidelity scaling [27] and generalizability [28] instead to harder *excited-state* NNQMD. We have achieved 1590.31 [sec]/(1,228,800,000,000 [atoms] × 690,000 [weights]) = $1.876 \times 10^{-15}$ [sec] on 10,000 Aurora nodes. This amounts to 3,780-fold reduction compared to the SOTA T2S of $7.091 \times 10^{-12}$ [sec] [11]; see Table II.

### VIII. IMPLICATIONS

As AI becomes ubiquitous at every corner of our society, its enormous power demand exposes fundamental physics limits like the Boltzmann tyranny [1]. Developing ultrafast and ultralow-power computing and sensing devices to overcome these limits will require new developments in attosecond physics and topological quantum matter (or topotronics). Disruptive possibilities include petahertz electronics [61], as well as topologically protected attojoule-switching logic and robust quantum computing [55, 62], which are many orders-of-magnitude faster and less energy-consuming than the current CMOS technology. However, this is a formidable multiscale/multiphysics problem involving multiple field and particle equations for light, electrons, and atoms, encompassing electron/atom-to-device scales. The 2024 Nobel physics and chemistry prizes heralded the new era, where AI is embedded into the very fabric of science to attack such challenges. This work is an exemplar, using AI-enhanced NNQMD to boost first-principles NAQMD. Meanwhile, HPC is at a historic crossroads, where increasingly heterogeneous computing hardware focusing on low-precision arithmetic puts traditional modeling and simulation applications at a risk of extinction [17]. This work unleashed the power of AI-enhanced multiscale/multiphysics simulation, where the divide-conquer-recombine/metamodel-space-algebra (DCR/MSA) paradigm with parameterized precision exploits increasingly heterogeneous exascale computing platforms that support a spectrum of hybrid precision modes. This new paradigm offers viable algorithm-hardware co-design pathways for a wide variety of multiscale/multiphysics problems in the coming post-exascale era. Moreover, our open parallel programming approach provides the benefit of post-exascale computing to the general HPC community. We are using our multiscale-light-matter-dynamics software on Aurora to inform SOTA X-ray free-electron laser experiments at the newly upgraded LCLS at Stanford [63]. Such integrated computational/experimental studies will be critical not only for future topotronics but also for a wide range of advanced technologies. Furthermore, DCR/MSA allows exponentially hard topological quantum many-body dynamics subproblems [64] to be offloaded from Exaflop/s supercomputers to emerging quantum processing units (QPUs) with minimal CPU-QPU communication, *i.e.*, realizing quantum-centric supercomputing [65]. Thus, the DCR/MSA paradigm is metascalable at the emerging nexus of post-exascale HPC, AI, and quantum computing.

ACKNOWLEDGMENT

This work was supported by U.S. DOE award DE-SC0023146. AN was supported by Office of Naval Research through a Multi-University Research Initiative (MURI) grant N00014-24-1-2313. KN was supported by NSF grant OAC-2118061. Computer time was provided by DOE Aurora ESP and INCITE programs. This research used resources from the Argonne Leadership Computing Facility, a DOE Office of Science user facility at Argonne National Laboratory, which is supported by DOE contract DE-AC02-06CH11357. Figure 3 was created by Joseph Insley, Victor Mateevisti, Silvio Rizzi, and Janet Knowles.